\documentclass[aps,prl,twocolumn,superscriptaddress,nofootinbib,longbibliography,floatfix]{revtex4-2}

\usepackage[dvipsnames]{xcolor}
\usepackage{amsmath,amssymb,booktabs,graphicx,bm,microtype,slashed}
\usepackage[colorlinks=true,citecolor=blue!55!black,linkcolor=blue!55!black,urlcolor=blue!55!black]{hyperref}

\newcommand{\kms}{\mathrm{km/s}}
\newcommand{\cmg}{\mathrm{cm^2/g}}

\newcommand{\mnras}{MNRAS}

\begin{document}

\title{Mass-Dependent Dark Matter Deficit from Inelastic Scattering}

\author{Daneng Yang }                                      \email{yangdn@pmo.ac.cn}

\author{Yi-Zhong Fan }
\email{yzfan@pmo.ac.cn}

\affiliation{$^1$Purple Mountain Observatory, Chinese Academy of Sciences, Nanjing 210033, China}
\affiliation{$^2$School of Astronomy and Space Sciences, University of Science and Technology of China, Hefei 230026, China} 

\date{\today}

\begin{abstract}
Recent stellar-kinematic and neutral-hydrogen observations indicate a dark matter deficit within the central tens of kiloparsecs of nearby galaxies that grows systematically with stellar mass relative to hydrodynamical simulations. We show that this mass dependence can arise from exothermic inelastic dark matter with strongly velocity-dependent scattering. Two nearly degenerate dark matter states interact through vector and scalar mediators with opposite-sign contributions and unequal ranges. The resulting coupled-channel dynamics suppresses $s$-wave conversion at low velocity while retaining a $p$-wave enhancement at several hundred $\kms$, where down-scattering injects kinetic energy and lowers central dark matter densities. At dwarf velocities, conversion remains suppressed while elastic scattering can still drive core formation consistent with the observed dwarf-clustering pattern. A late dark-sector phase transition, along with the low-velocity suppression, preserves a large excited-state fraction until halo formation. Using representative halos spanning the four observed stellar-mass bins, we find that the model accounts for the inferred dark matter deficit in all four bins, including its systematic growth with stellar mass, while its impact weakens toward cluster velocities. The growing discrepancy with hydrodynamical simulations may therefore offer a glimpse of previously hidden dark matter microphysics. 
\end{abstract}

\maketitle

{\noindent\bf Introduction.} Dark matter constitutes about 85\% of the Universe's matter content and provides the gravitational potential wells in which galaxies form and evolve~\cite{Planck:2018vyg}. Recent observations, however, reveal an unexpected systematic trend in the dark matter distributions of nearby galaxies. Combining spatially resolved stellar kinematics with neutral hydrogen information for 136 galaxies, Lei \textit{et al.} ({\it Nature Astronomy} in press) infer dark matter densities increasingly below the predictions of the TNG100~\cite{2018MNRAS.475..676S}, EAGLE~\cite{2015MNRAS.446..521S}, SIMBA~\cite{Dave:2019yyq}, and AIDA-TNG~\cite{Despali:2025koj} simulations toward larger stellar masses~\cite{Lei2026}. The deficit extends from approximately $10~\mathrm{kpc}$ to more than $50~\mathrm{kpc}$ toward the most massive systems. An independent analysis combining MaNGA~\cite{2015ApJ...798....7B}, ALFALFA~\cite{Giovanelli:2005ee}, and group-calibrated halo masses finds the same trend, with an approximately $4\sigma$ population-level separation at the high-mass end~\cite{Wang2026}. These results reveal a dark matter deficit that becomes progressively more pronounced with increasing galaxy mass.

The origin of this mass dependence is not obvious within standard galaxy formation. Stellar feedback becomes less effective toward Milky Way and larger halo masses, where baryonic condensation can instead contract the dark matter distribution~\cite{DiCintio:2013qxa,Chan:2015tna,2016MNRAS.456.3542T,2020MNRAS.497.2393L}. At the massive end, AGN feedback, mergers, and dynamical heating can naturally produce halo expansion and may account for part of the observed deficit~\cite{Lei2026,Wang2026}. The persistence of the deficit across a broad mass range, alongside independent constraints at cluster scales, motivates exploring dark matter physics whose efficiency varies with halo velocity scale, as naturally occurs for velocity-dependent interactions~\cite{Loeb:2010gj,2012MNRAS.423.3740V,2013MNRAS.431L..20Z,Tulin:2013teo,Kaplinghat:2015aga,Yang:2022mxl,Nadler:2023nrd,Ohana:2026bwl,Nadler:2026waz}.

We show in this Letter that the inelastic dark matter ~\cite{Tucker-Smith:2001myb,Arkani-Hamed:2008hhe,Cui:2009xq,Graham:2010ca,Schutz:2014nka,An:2020tcg,Vogelsberger:2018bok,Duan:2026zqj} can account for this mass dependence through exothermic scattering that heats and evaporates dark matter from halo centers. We consider two nearly degenerate dark matter states, $\chi_1$ and $\chi_2$, with $m_2-m_1\equiv\delta>0$. The exothermic process $\chi_2\chi_2\to\chi_1\chi_1$ converts the mass splitting into kinetic energy, characterized by the kick velocity $v_k=\sqrt{2\delta/m_\chi}$. 
For an incoming relative velocity $v_r$, energy conservation gives $v_{r,22\to11}'=\sqrt{v_r^2+4v_{\rm k}^2}$, and
$v_{r,11\to22}'=\sqrt{v_r^2-4v_{\rm k}^2}$.
The inverse process is forbidden for $v_r<2v_{\rm k}$, introducing a characteristic velocity scale that, together with the velocity dependence of the scattering amplitude, can localize the effect to galaxy-scale halos.

{\noindent\bf Model and initial conditions.} The observed mass dependence imposes two simultaneous requirements. Down-scattering must be weak at the characteristic velocities of lower-mass galaxies, where no comparable dark matter deficit is observed, while becoming efficient at several hundred $\kms$ in massive halos. At the same time, a significant excited-state population must survive until halo formation to supply the exothermic energy. The same low-velocity suppression is also essential cosmologically. Alongside a late dark-sector phase transition that generates the mass splitting, it leaves the excited-state reservoir nearly intact until halo formation. At still higher velocities, the interaction must remain consistent with cluster-scale constraints.

\begin{figure}[t]
 \centering
 \includegraphics[width=\columnwidth]{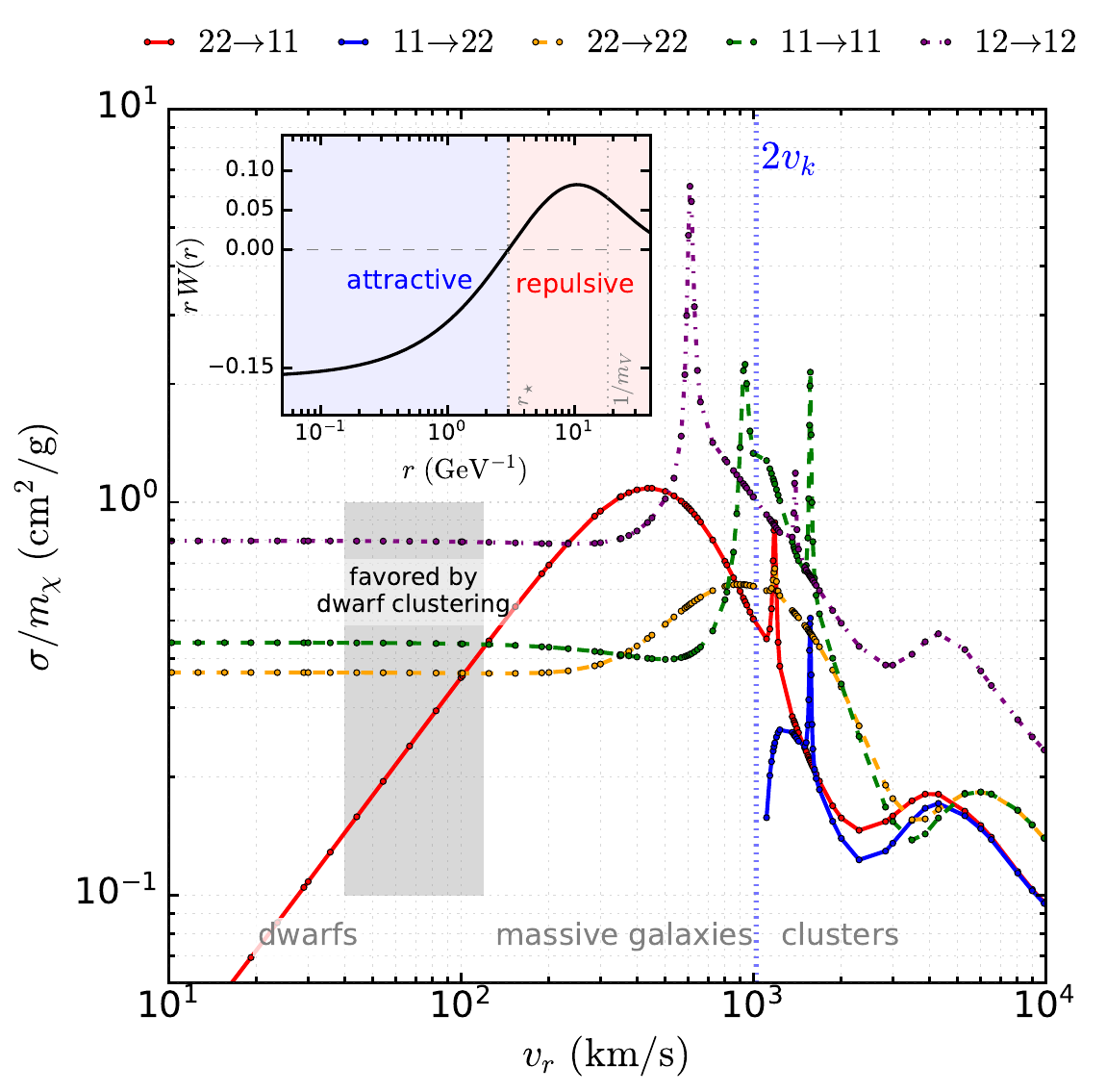}
 \caption{Five-channel cross sections per unit dark matter mass for the benchmark in Eq.~\eqref{eq:benchmark}. Down-scattering is strongly suppressed at dwarf velocities and enhanced at massive-galaxy velocities, where the $p$ wave dominates, before falling at cluster velocities. The inverse $11\to22$ channel opens at $v_r=2v_k$ (dotted line). Inset: the dimensionless potential $r\,W(r)$, attractive inside the node $r_\star$ and repulsive outside. The gray shaded region indicates the area preferred by the observed clustering of dwarf galaxies~\cite{Zhang:2025bju}.
}
 \label{fig:rates}
\end{figure}

For finite-range exothermic scattering, the generic Wigner threshold law gives
$\sigma^{(\ell)}\propto v_r^{2\ell-1}$, so an unsuppressed $s$ wave scales as
$v_r^{-1}$, while the $p$ wave scales as $v_r$~\cite{Wigner:1948zz,Tomza:2019woz,Sadeghpour_2000,Cassel:2009wt}.
We focus on the resonant regime, where nonperturbative partial-wave structure can enhance scattering at galaxy velocities while preserving suppressed rates at lower and higher velocities.
A single Yukawa interaction ties this low-velocity $s$ wave and the galaxy-scale $p$-wave enhancement to the same force range, making the required velocity hierarchy difficult to realize. We instead combine two finite-range interactions with opposite signs and unequal ranges,
\begin{equation}
\label{eq:Wr}
W(r)=\frac{\alpha_V e^{-m_V r}-\alpha_S e^{-m_S r}}{r},
\end{equation}
which changes sign at
$r_\star=\ln(\alpha_V/\alpha_S)/(m_V-m_S)$,
being attractive at short distances and repulsive at large distances, as illustrated in the inset of Fig.~\ref{fig:rates}. The resulting radial structure drives the $s$-wave conversion amplitude through a zero while retaining the $p$-wave enhancement at galaxy velocities.
A minimal realization consists of two nearly degenerate Dirac flavor
states coupled with opposite charges to a vector and a scalar mediator,
\begin{align}
 \mathcal L_{\rm mass}&= - m_\chi\bar\Psi\Psi -
 \left(y_\Phi\Phi\,\bar\psi_+\psi_-+\mathrm{H.c.}\right),
 \label{eq:pseudodiracmass}\\
 \mathcal L_{\rm int}&=
 g_VV_\mu\bar\Psi\gamma^\mu\sigma_3\Psi
 +g_SS\bar\Psi\sigma_3\Psi ,
 \label{eq:lagrangian}
\end{align}
where $\Psi=(\psi_+,\psi_-)^T$. 
A late expectation value of $\Phi$ mixes the flavor states, producing mass eigenstates $\chi_1=(\psi_+-\psi_-)/\sqrt{2}$ and $\chi_2=-i(\psi_++\psi_-)/\sqrt{2}$, with splitting $\delta=m_2-m_1=2y_\Phi\langle\Phi\rangle$. 
Each flavor state is therefore an equal superposition of $\chi_1$ and $\chi_2$, so decoherence after the transition naturally yields $f_2\simeq1/2$ as the initial condition for halo evolution.

The relic abundance can be established before the transition through an asymmetric Dirac population~\cite{Kaplan:2009ag,Petraki:2011mv,Zurek:2013wia,Tulin:2012re,Yang:2025dgl}, while late changes in dark matter properties after chemical freeze-out have several known realizations~\cite{Cohen:2008nb,Coleman:1977py,Callan:1977pt,Garny:2024ums,Balan:2025uke,Guo:2026cuv,Huang:2026qdc}.  We assume that the mediators decay before nucleosynthesis, for instance through kinetic or Higgs mixing~\cite{Holdom:1985ag,Patt:2006fw,Fradette:2014sza}. We avoid radiative bound-state formation by requiring on-shell mediator emission to be kinematically forbidden~\cite{Wise:2014jva}, which sets an upper boundary of the viable parameter region.

After rotating to the mass basis, both mediator interactions are off-diagonal. Vector exchange is repulsive between dark particles, whereas scalar exchange is attractive, and the two-body operator is $V_{\rm NR}(r)=W(r)\sigma_2^{(1)}\sigma_2^{(2)}$, with $|1\rangle\equiv|\chi_1\rangle$, $|2\rangle\equiv|\chi_2\rangle$, and $|ij\rangle\equiv|i\rangle\otimes|j\rangle$, one has
$\sigma_2^{(1)}\sigma_2^{(2)}|11\rangle=-|22\rangle$ and
$\sigma_2^{(1)}\sigma_2^{(2)}|12\rangle=+|21\rangle$.
The resulting two-state dynamics is therefore described by~\cite{Slatyer:2009vg}
\begin{align}
 V_{11,22}(r)&=
 \begin{pmatrix}
 0&-W(r)\\
 -W(r)&2\delta
 \end{pmatrix},
 \nonumber\\
 V_{12,21}(r)-\delta\mathbf1&=
 \begin{pmatrix}
 0&+W(r)\\
 +W(r)&0
 \end{pmatrix}.
 \label{eq:potentialblocks}
\end{align}
The relative sign between these sectors follows directly from the flavor operator. The unequal mediator ranges then generate nontrivial partial-wave structure~\cite{Tulin:2013teo,Schutz:2014nka}. For the parameter region of interest, the $s$-wave conversion amplitude develops a zero near dwarf velocities, whereas a $p$-wave resonance enhances down-scattering at massive-galaxy velocities. The same splitting sets the exothermic kick velocity $v_k=\sqrt{2\delta/m_\chi}$ and opens the inverse process $\chi_1\chi_1\rightarrow\chi_2\chi_2$ only for $v_r>2v_k$. The model therefore suppresses energy injection in low-mass galaxies, enhances it in massive galaxies, and activates endothermic conversion toward cluster velocities.

We obtain the scattering amplitudes by integrating the coupled radial Schr\"odinger equation and constructing the flux-normalized $S$ matrix. For identical unpolarized spin-$1/2$ particles in the $11$ and $22$ channels,
\begin{equation}
 \sigma_{i\to j}=
 \frac{\pi}{k_i^2}
 \sum_\ell
 2w_\ell(2\ell+1)
 \left|S_{ji}^{(\ell)}-\delta_{ji}\right|^2 ,
 \label{eq:sigmaidentical}
\end{equation}
where $w_\ell=1/4$ for even singlet waves and $3/4$ for odd triplet waves. The mixed $12$ channel is treated as distinguishable. Detailed balance gives
$v_{r,\rm up}^2\sigma_{11\to22}=v_{r,\rm down}^2\sigma_{22\to11}$
with $v_{r,\rm down}=\sqrt{v_{r,\rm up}^2-4v_{\rm k}^2}$.

\begin{figure*}[t]
 \centering
 \includegraphics[width=\textwidth]{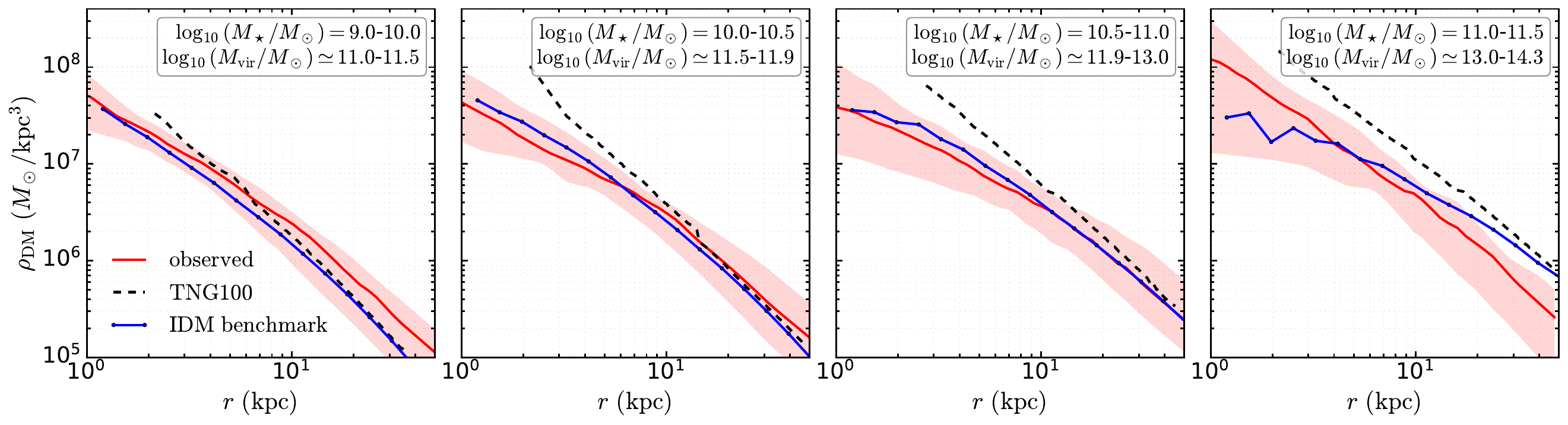}
 \caption{Dark matter density profiles at an exact $10\,\mathrm{Gyr}$ snapshot with $10^6$ dark matter particles per bin. Blue curves show the inelastic model, red curves and bands show the observed median and $1\sigma$ population scatter, and dashed curves show TNG100 digitized from Ref.~\cite{Lei2026}. Text boxes give the stellar-mass bins and indicative halo-mass ranges from the $z=0$ median stellar-to-halo relation of Ref.~\cite{2013MNRAS.428.3121M}. At this binning the sampling noise of a single snapshot stays below $8\%$ outside $5\,\mathrm{kpc}$ and rises to about $40\%$ at the innermost resolved point of the most massive bin. The same microscopic interaction is used in all four bins.}
 \label{fig:galaxies}
\end{figure*}

A representative benchmark is
\begin{equation}
\begin{split}
 m_\chi &=12.4\,\mathrm{GeV},
 \qquad
 \delta=18.1\,\mathrm{keV},\\
 m_V&=54.8\,\mathrm{MeV},
 \qquad
 \alpha_V=0.184,\\
 m_S&=266\,\mathrm{MeV},
 \qquad
 \alpha_S=0.347 .
\end{split}
\label{eq:benchmark}
\end{equation}
It gives $v_k=513\,\kms$ and $\sigma_{22\to11}(300~\kms)/m_\chi=0.949~\cmg$. Down-scattering is broadly enhanced around $500~\kms$, where the $p$ wave dominates, while the low-velocity rate is strongly suppressed, with
$K_{22\to11}(30~\kms)/K_{22\to11}(300~\kms)=0.0114$ and
$K\equiv(\sigma/m_\chi)v_r$.
The inverse channel opens at $2v_k=1026\,\kms$, naturally placing its threshold at cluster velocities.
Figure~\ref{fig:rates} shows the velocity dependence of all five scattering channels. The solid red curve denotes the exothermic process $\chi_2\chi_2\to\chi_1\chi_1$, which peaks near $500\,\kms$ and drives the dark matter deficit in massive galaxies. At lower velocities, the elastic channels lie within the gray shaded region favored by the dwarf-clustering observation of Ref.~\cite{Zhang:2025bju}.

The late phase transition prepares an excited-state fraction
$f_{2,\rm PT}\simeq1/2$, while the low-velocity suppression of
down-scattering prevents this reservoir from being significantly depleted
before halo formation. Defining
$K_{ij}\equiv(\sigma_{ij}/m_\chi)v_r$, its homogeneous evolution obeys
\begin{equation}
 \dot f_2= \rho_\chi
 \left[
 (1-f_2)^2\langle K_{11\to22}\rangle - f_2^2\langle K_{22\to11}\rangle
 \right].
 \label{eq:f2boltzmann}
\end{equation}
For the benchmark above, the low-velocity rate approaches
$K_{22\to11}\to K_{0,\rm b}=0.0135\,(\cmg)\,\kms$.
In the cold-background limit, up-scattering is negligible and the
depletion is controlled by the integrated conversion depth
\begin{equation}
\tau_{\rm conv}\equiv\int_{t_{\rm PT}}^{t_f}\rho_\chi K_0\,dt
\simeq 4.36\times10^{-4}
\frac{K_0}{K_{0,\rm b}}
\left(\theta_{\rm PT}^{3/2}-\theta_f^{3/2}\right),
\label{eq:tauconv}
\end{equation}
where $\theta\equiv T_\gamma/\mathrm{eV}$. The surviving fraction, $f_2(z_f)\simeq[f_{2,\rm PT}^{-1}+\tau_{\rm conv}]^{-1}$, therefore remains close to its initial value for a sufficiently late transition because the low-velocity floor keeps $\tau_{\rm conv}\ll1$.

Down-scattering nevertheless releases $2\delta$ per conversion, heating the dark sector and enhancing the conversion rate. We bracket this feedback between the cold limit above and instantaneous energy sharing, $\dot T_\chi+2HT_\chi=-(2\delta/3)\dot f_2$. Because the heating term is proportional to $\dot f_2$, it only amplifies depletion already generated at the low-velocity floor and remains negligible for halo evolution as long as the excited reservoir survives. For $T_{\gamma\,\rm PT}=0.8\,\mathrm{eV}$, instantaneous sharing reduces $f_2(z=20)$ from $0.500$ to $0.495$. The same feedback constrains how early the transition can occur, since it is seeded by the depletion and grows with it. Taking $f_2(z=20)=0.4$ as a deliberately loose criterion, instantaneous sharing requires $T_{\gamma,\rm PT}$ to be of order $1\,\mathrm{eV}$, whereas the cold limit allows values up to about $8\times10^{2}\,\mathrm{eV}$. The true bound therefore lies between the eV and keV scales, with later transitions bringing $f_2$ progressively closer to one half.

{\noindent\bf Halo simulations.}
We study the impact of this inelastic dark matter model on stellar-halo systems using a simplified $N$-body framework. We evolve isolated spherical halos with live dark matter and stellar particles, reducing the six-dimensional phase space to radius and speed following Refs.~\cite{Kamionkowski:2025uae,Gurian:2025zpc}.
Both components are initialized as Dehnen profiles~\cite{Dehnen:1993uh} fitted to the digitized TNG100 dark matter distributions and the observed stellar profiles of Ref.~\cite{Lei2026}, with equilibrium distribution functions obtained by Eddington inversion in the combined potential. Particles are paired locally and scattered using the tabulated multichannel rates and angular distributions, while conversion events enforce the inelastic kinematics. Particles with positive orbital energy remain in the calculation, allowing halo expansion and evaporation to emerge dynamically. 
Further details of the initial conditions, binning, and numerical tests are given in the Supplemental Material.

Figure~\ref{fig:galaxies} shows the resulting profiles. The same microscopic parameters are used for all four stellar-mass bins. The interaction has little effect in the lowest-mass systems because the down-scattering rate is suppressed at their characteristic velocities. The rate rises across the intermediate and high-mass bins, while the fixed kick speed provides an additional dependence through $v_{\rm k}/v_{\rm esc}$. The model consequently produces a dark matter deficit that becomes more prominent and extends to larger radii toward higher galaxy mass.
The simulated profiles lie within the observed population scatter over most of the radial range in all four stellar-mass bins, reproducing both the magnitude and the systematic mass dependence of the inferred dark matter deficit. The published bands represent galaxy-to-galaxy scatter, so the comparison is interpreted at the population level.

To test the suppression at cluster velocities, we evolve the M15B cluster halo with its live Hernquist stellar component from Ref.~\cite{Yang:2025xsp} under the same benchmark model. The final density profile closely follows the elastic-only result and is comparable to that obtained for elastic self-interactions with $\sigma/m=0.3\,\mathrm{cm^2/g}$. Reverse conversion offsets most of the exothermic heating, although elastic heat transport remains important and the net conversion effect remains weakly exothermic.

Several cluster studies have reported dark matter profiles substantially shallower than NFW, including core-like profiles on scales of $\mathcal{O}(10)\,\mathrm{kpc}$ inferred from lensing and BCG kinematics~\cite{2013ApJ...765...25N,Sagunski:2020spe,Cerny:2025zfn}, while other analyses favor smaller cores and constrain constant self-interactions to $\sigma/m\lesssim0.1$--$0.3\,\cmg$~\cite{Sagunski:2020spe,Andrade:2020lqq,Eckert:2022qia}. The cluster-scale situation therefore remains observationally open. Our benchmark predicts a measurable core, generated primarily by elastic heat transport after the up- and down-scattering channels partially cancel. Improved measurements of central cluster mass profiles thus provide a direct test of the same interaction responsible for the galaxy-scale deficit.

\begin{figure}[t]
\centering
\includegraphics[width=\columnwidth]{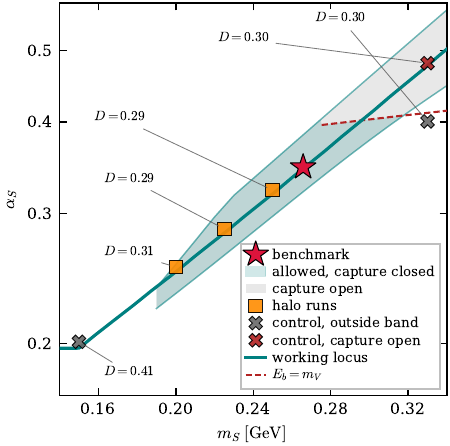}
\caption{Viable region in the $(m_S,\alpha_S)$ plane with the remaining parameters fixed to Eq.~\eqref{eq:benchmark}. The teal curve marks the relevant $s$-wave conversion zero, while the shaded band additionally satisfies the low- and galaxy-velocity rate requirements. The star denotes the benchmark and orange squares show additional points tested with full halo simulations, labeled by the profile discrepancy $D$ defined in the text. Control points illustrate the loss of the galaxy-scale enhancement or excessive low-velocity conversion outside the band. The red dashed contour, $E_b=m_V$, marks the onset of radiative bound-state formation and sets the upper boundary of the viable region.
}
\label{fig:locus}
\end{figure}

{\noindent\bf Model parameter space.}
The viable region is organized around the zero of the $s$-wave conversion amplitude. Figure~\ref{fig:locus} shows the $(m_S,\alpha_S)$ plane with $m_\chi$, $\delta$, $m_V$, and $\alpha_V$ fixed to the benchmark values. The condition $|S_{21}^{(\ell=0)}|=0$ defines the teal locus. The phenomenologically useful region has finite width because the scattering rates need only satisfy
$\sigma_{22\to11}/m_\chi<0.2\,\cmg$ at $30\,\kms$ and $\sigma_{22\to11}/m_\chi>0.5\,\cmg$ at $300\,\kms$.
These requirements select the shaded band around the locus, with a width of about $0.08$ dex in $\alpha_S$ near the benchmark.

The orange squares mark three further points evolved with full halo simulations.  To compare them against the benchmark we use the discrepancy $D$, defined as the rms logarithmic difference between the simulated and observed dark matter densities evaluated in three equally weighted shells at $3-5$, $5-10$, and $10-20\,\mathrm{kpc}$ and then averaged with equal weight over the four stellar-mass bins.  The benchmark gives $D=0.28$ and the three additional points give $D=0.29-0.31$, so the agreement is not special to one parameter choice.  The band closes toward low $m_S$ as the galaxy-scale enhancement becomes too weak, illustrated by the control at $m_S=0.15\,\mathrm{GeV}$ where $D$ degrades to $0.41$.  Moving off the $s$-wave zero restores excessive low-velocity conversion instead, as shown by the control displaced from it by $\Delta\alpha_S\simeq0.08$ at $m_S=0.33\,\mathrm{GeV}$. At still larger coupling, radiative bound-state formation becomes kinematically allowed once the deepest binding energy exceeds $m_V$. We therefore truncate the viable region at the red dashed contour $E_b=m_V$. The working parameter space is thus a finite band along one scattering-matrix branch, rather than an isolated tuned point.

{\noindent\bf Discussion and conclusions.} We have shown that a two-state dark sector produced by a late phase transition can reproduce the observed mass-dependent dark matter deficit through velocity-dependent scattering. Conversion is suppressed at dwarf velocities, becomes strongly exothermic in massive galaxies, and weakens again toward clusters as the inverse process opens, while elastic scattering remains active at the low-velocity end. The same benchmark matches the inferred dark matter profiles in all four stellar-mass bins and predicts correlated signatures from dwarfs to clusters.

Baryonic processes such as AGN feedback and mergers may contribute to the inferred deficit~\cite{Lei2026,Wang2026}, but whether they can reproduce its systematic growth across the observed mass range remains to be shown. In the inelastic interpretation, this mass dependence follows directly from the velocity dependence of the scattering dynamics. Improved galaxy and cluster mass profiles can therefore test whether the observed trend reflects baryonic evolution or a correlated sequence of dark matter interactions across halo velocity scales.

\begin{acknowledgments}
The authors used generative AI tools for code-assistance and language editing during software and manuscript development. 
The scientific content, results, and conclusions were designed, verified, and approved by the authors.
\end{acknowledgments}

\bibliography{references}

%\clearpage
\appendix

\section{Cluster scale constraints}
\label{app:cluster}

\begin{figure}[t]
 \centering
 \includegraphics[width=\columnwidth]{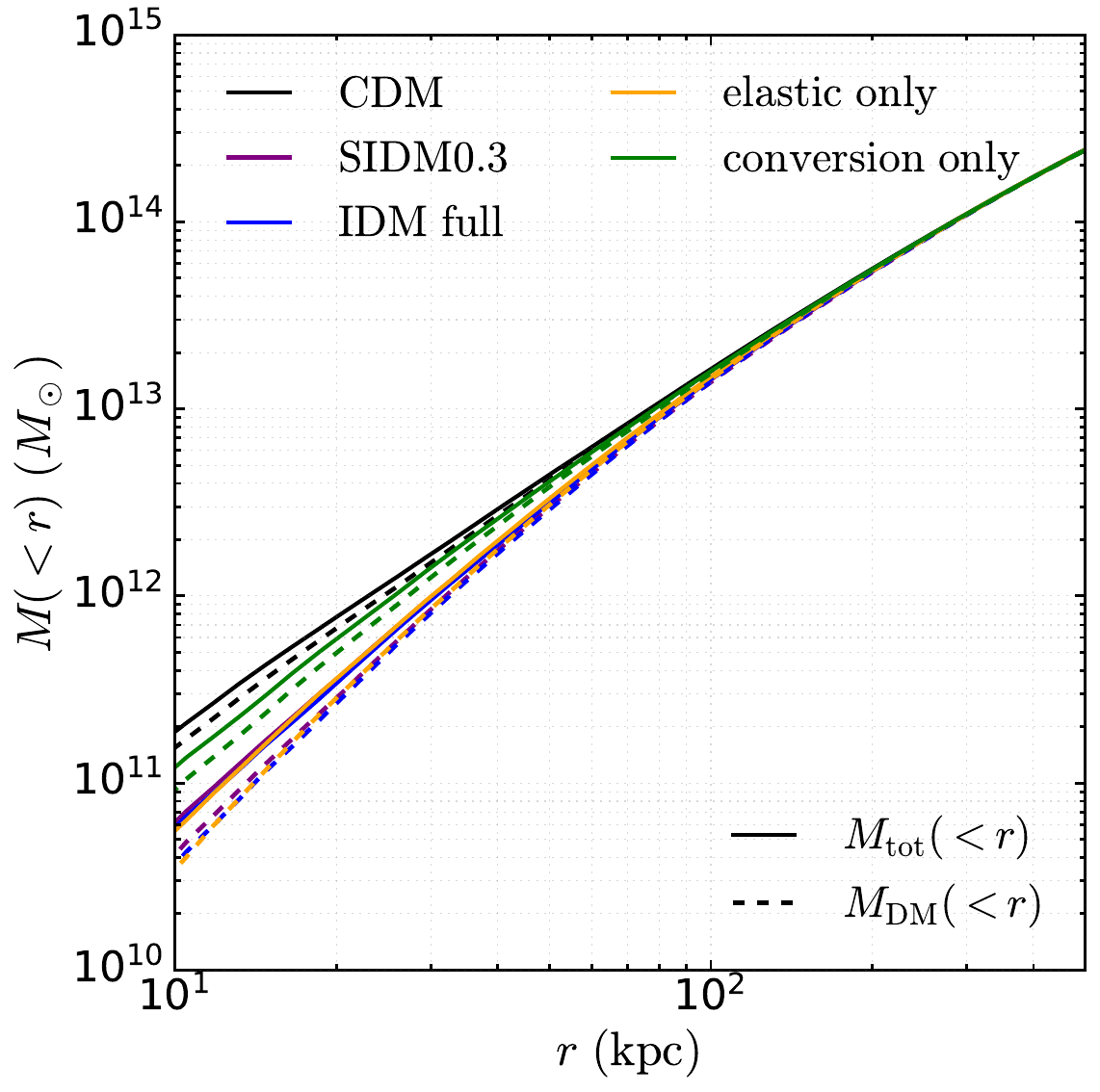}
 \caption{Enclosed mass profiles of the M15B cluster after $5\,\mathrm{Gyr}$ with live stellar baryons. Every case appears twice, as total enclosed mass (solid) and as dark matter mass (dashed). Endothermic conversion removes most of the exothermic conversion imbalance, while elastic transport and the residual net down-scattering leave an appreciable core. The full model closely tracks the constant $\sigma/m=0.3,\cmg$ elastic control outside the central region, with only small differences beyond $\sim50,\mathrm{kpc}$.}
 \label{fig:cluster}
\end{figure}

The cluster-scale continuation is tested with the M15B halo of Ref.~\cite{Yang:2025xsp}, using a live Hernquist stellar component with $M_\star=7.5\times10^{11}M_\odot$ against $M_{\rm DM}=2.4\times10^{15}M_\odot$, $10^6$ dark matter and $311$ stellar particles, $5.4\,\mathrm{kpc}$ gravitational softening, and a $5\,\mathrm{Gyr}$ evolution. The run records $5275$ down-scattering and $4177$ up-scattering events. Up-scattering thus offsets $79\%$ of the conversion events in the exothermic direction, leaving a residual net conversion that still heats the halo.

Figure~\ref{fig:cluster} shows the resulting enclosed mass profiles. Against a matched run with constant elastic $\sigma/m=0.3\,\cmg$, the full model agrees to within a few percent over the same radial range, so at cluster scale the inelastic benchmark is not separable from an elastic control of that size.
The same figure separates the two contributions. The full result is close to the elastic-only case, while removing elastic scattering greatly reduces the halo response. Cluster-scale evolution is therefore driven mainly by elastic heat transport, with incomplete cancellation between up- and down-scattering providing a smaller additional effect.

\section{Coupled-channel scattering calculation and validation}
\label{app:coupledchannel}

This appendix presents the coupled-channel scattering equations used to construct the five velocity- and angle-dependent kernels in the halo simulations. We then validate the implementation against published one- and two-channel results.

\subsection{Radial equation and asymptotic matching}

For each total orbital angular momentum $\ell$, the reduced radial matrix $\bm U_\ell$ obeys
\begin{equation}
 \bm U_\ell''(r)+\left[\bm k^2-
 \frac{\ell(\ell+1)}{r^2}\bm 1-2\mu\bm V(r)\right]
 \bm U_\ell(r)=0,
 \label{eq:app_radial}
\end{equation}
where $\mu\simeq m_\chi/2$ and the diagonal matrix $\bm k$ contains the open-channel momenta at the chosen center-of-mass energy. A channel below threshold instead has $k_i=i\kappa_i$ and is matched to the decaying modified spherical-Bessel solution. 
We propagate a regular matrix solution from the origin, periodically orthogonalizing it to control exponentially growing closed-channel components, and form its logarithmic derivative $\bm Y=\bm U_\ell'\bm U_\ell^{-1}$. Matching $\bm Y$ to the free incoming and outgoing solutions at a radius where both Yukawa terms are negligible gives the flux-normalized $S$ matrix. The physical cross sections are then assembled with Eq.~\eqref{eq:sigmaidentical}. The distinguishable $12$ sector is summed with singlet and triplet weights but without an identical-final-state factor.

For a transition $i\to j$ we construct
\begin{equation}
 f_{ji}(\theta)=\frac{1}{2i\sqrt{k_i k_j}}
 \sum_{\ell=0}^{\ell_{\max}}(2\ell+1)
 \left(S_{ji}^{(\ell)}-\delta_{ji}\right)P_\ell(\cos\theta),
 \label{eq:app_amplitude}
\end{equation}
with the appropriate identical-fermion projection for the $11$ and $22$ sectors. The normalized cumulative distribution derived from $d\sigma_{i\to j}/d\Omega=|f_{ji}|^2$ is stored with the rate table and sampled by the halo code. Partial waves are increased until both the integrated rate and the angular distribution are stable. The radial range and the integration tolerance are checked independently at representative resonances, antiresonances, and thresholds. Convergence requires $\ell_{\max}\simeq25$,
and every number quoted below is computed with $\ell_{\max}=60$.

\subsection{Validation against published limits}

\begin{figure*}[t]
 \centering
 \begin{minipage}[t]{0.48\textwidth}
  \centering
  \includegraphics[width=\linewidth]{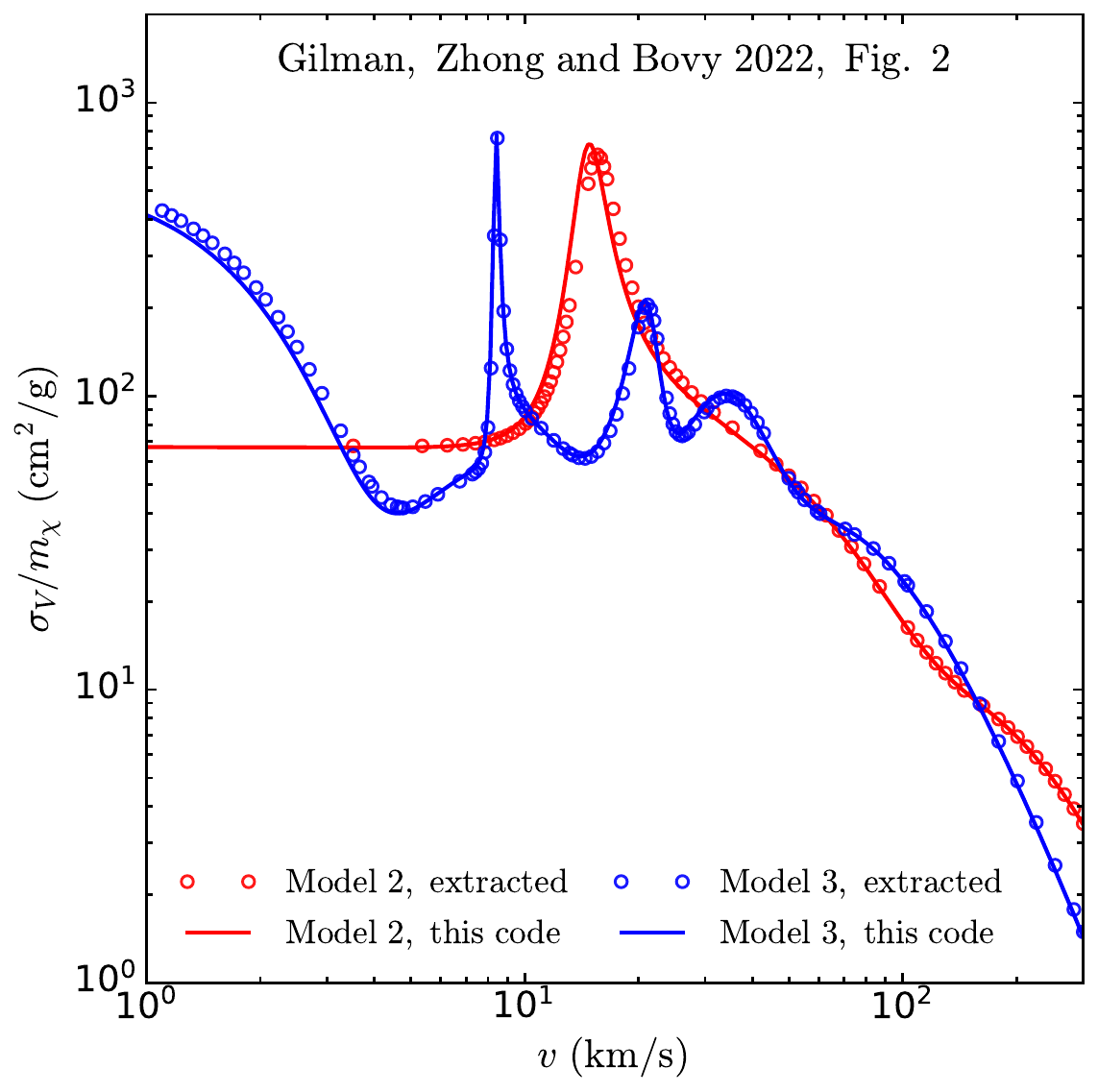}
 \end{minipage}\hfill
 \begin{minipage}[t]{0.48\textwidth}
  \centering
  \includegraphics[width=\linewidth]{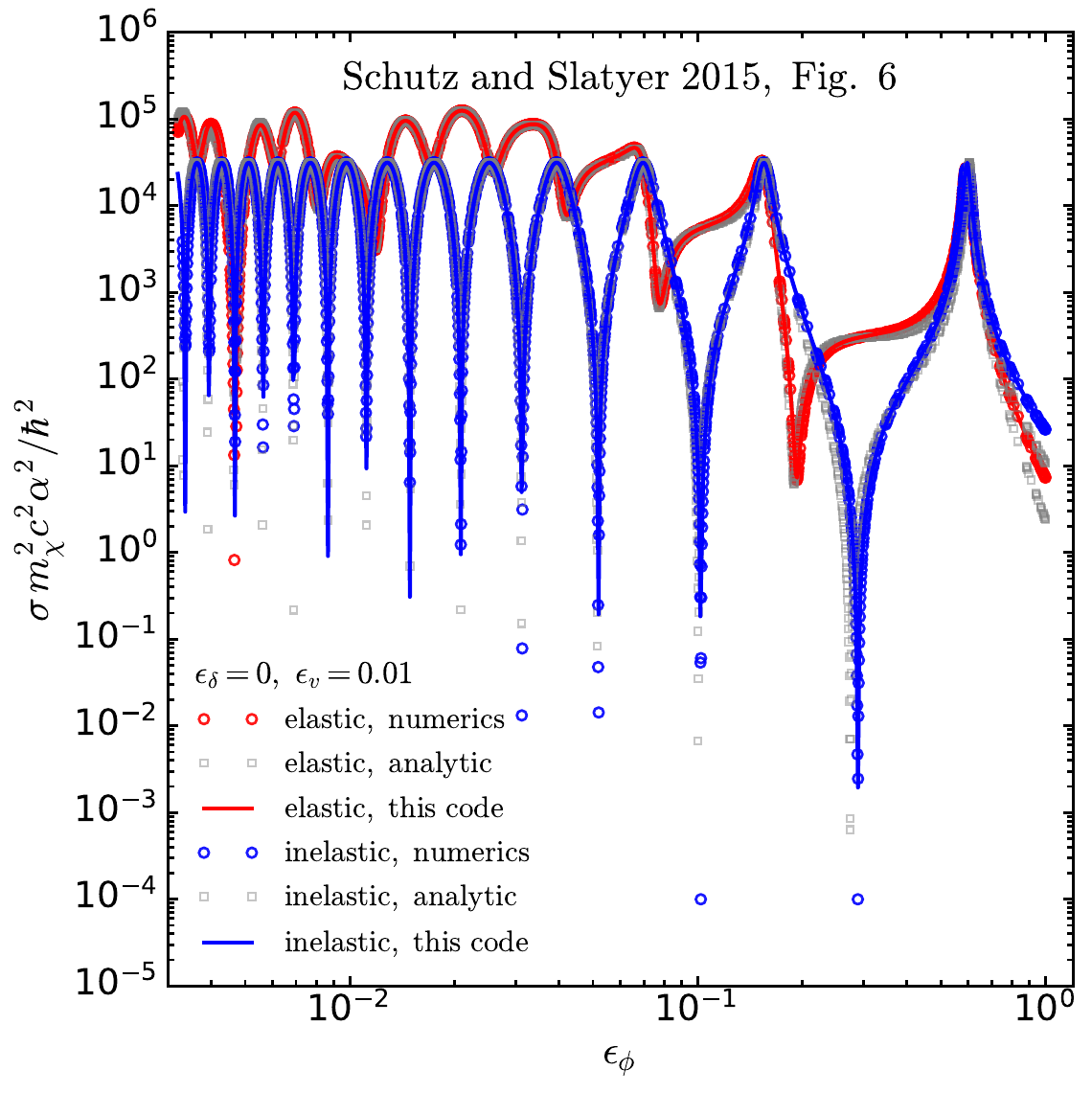}
 \end{minipage}
 \caption{Validation of the scattering solver in open channels. Left: single-channel viscosity cross sections for two resonant Yukawa models compared with Fig.~2 of Ref.~\cite{Gilman:2022ida}. Right: degenerate two-channel elastic and inelastic cross sections compared with Fig.~6 of Ref.~\cite{Schutz:2014nka}. Solid curves show the present calculation and symbols denote values extracted from the published results.}
 \label{fig:solver_validation_12}
\end{figure*}

Figure~\ref{fig:solver_validation_12} shows two complementary tests. In each panel the solid curves are calculated by the solver used in this work, while the open symbols are extracted from the corresponding published figure. The single-channel limit uses an attractive Yukawa potential and the viscosity cross section of Gilman, Zhong, and Bovy~\cite{Gilman:2022ida}. 
The calculation agrees well with the published result overall. The narrow low-velocity resonances are recovered after percent-level adjustments of the reported mediator parameters, validating the radial evolution, partial-wave sum, and unit conversion.

\begin{figure*}[t]
 \centering
 \begin{minipage}[t]{0.48\textwidth}
  \centering
  \includegraphics[width=\linewidth]{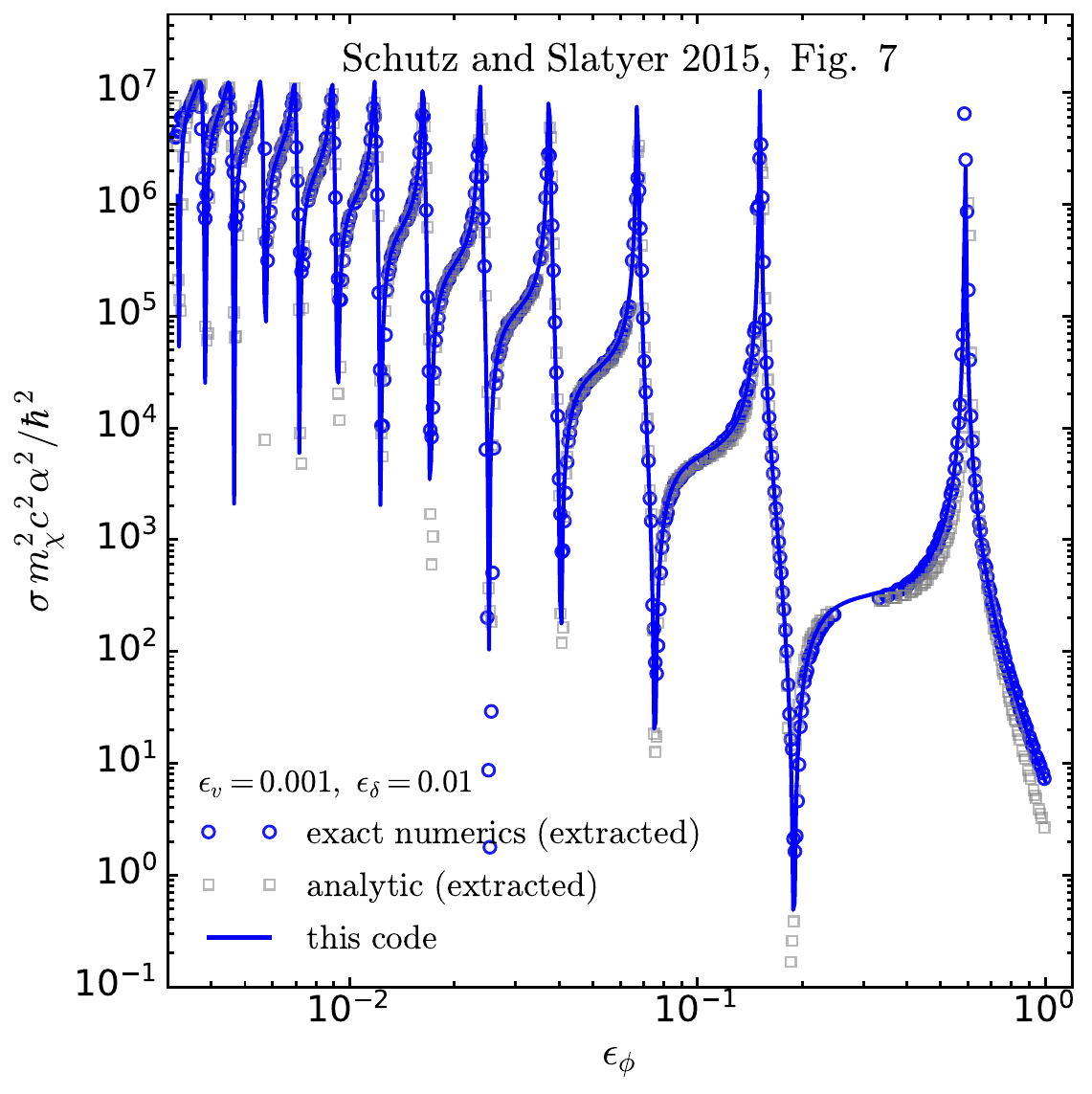}
 \end{minipage}\hfill
 \begin{minipage}[t]{0.48\textwidth}
  \centering
  \includegraphics[width=\linewidth]{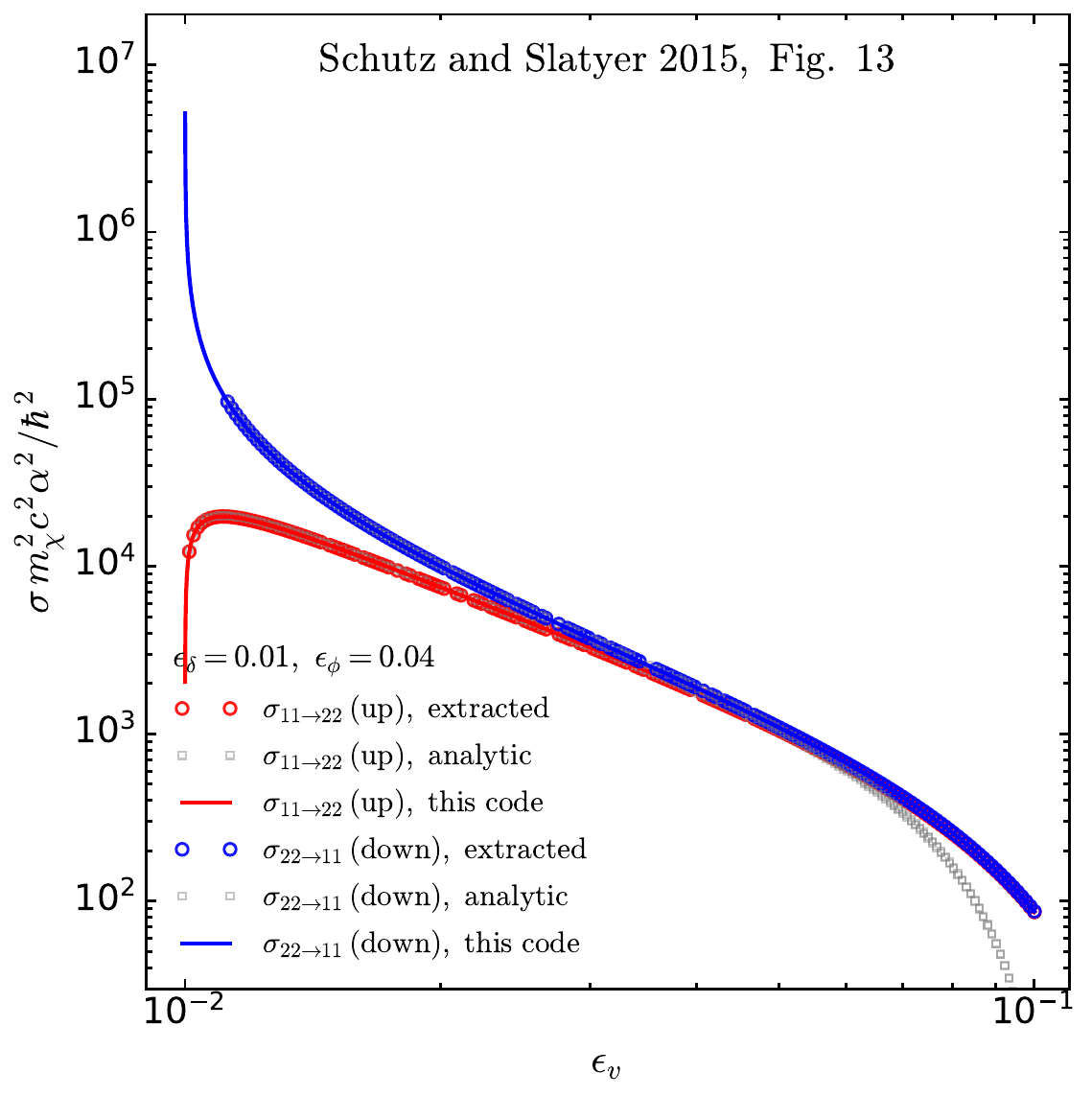}
 \end{minipage}
 \caption{Validation at finite mass splitting. Left: ground-state elastic scattering with the excited channel closed, compared with Fig.~7 of Ref.~\cite{Schutz:2014nka}. Right: endothermic and exothermic scattering above threshold, compared with Fig.~13 of the same reference. Solid curves show the present calculation and symbols denote values extracted from the published numerical and analytic results.}
 \label{fig:solver_validation_34}
\end{figure*}

The coupled-channel solver reproduces the benchmark results of Schutz and Slatyer~\cite{Schutz:2014nka} in both the degenerate and finite-splitting limits. Figure~\ref{fig:solver_validation_12} shows the degenerate test, including elastic and inelastic scattering, with agreement at the level expected from digitization.

Finite splitting provides additional tests of channel thresholds and inverse reactions. The left panel of Fig.~\ref{fig:solver_validation_34} tests the closed-channel boundary condition against Fig.~7 of Ref.~\cite{Schutz:2014nka}, while the right panel compares the open-channel results with their Fig.~13. For the latter, independent calculations of the forward and inverse reactions satisfy
\begin{equation}
v_{r,{\rm up}}^2\sigma_{11\to22} = v_{r,{\rm down}}^2\sigma_{22\to11},
\label{eq:app_detailed_balance}
\end{equation}
to machine precision and reproduce the expected Wigner threshold behavior~\cite{Wigner:1948zz}. These comparisons validate the channel thresholds, flux normalization, partial-wave treatment, and detailed balance used to construct the five scattering channels in the halo simulations.

\section{Spherical multichannel halo calculation and validation}
\label{app:halo}

This appendix summarizes the dynamical calculation used in the main text and validates its elastic limit against a standard self-interacting dark matter halo. The method follows the spherical phase-space reduction of Ref.~\cite{Kamionkowski:2025uae}, extended here to multichannel elastic and inelastic scattering. It retains velocity- and angle-dependent collision rates, state conversion, recoil, evaporation, and live stellar gravity while replacing the full three-dimensional force calculation by the spherical monopole. 

\subsection{Gravity and initial conditions}
\label{app:gravity}

Each particle carries $\bm{x}_i$, $\bm{v}_i$, and macro-particle mass $m_p$, with dark matter particles additionally carrying an internal-state label $s_i=1,2$. Gravity is computed from the instantaneous enclosed mass of the dark matter and stellar components,
\begin{equation}
\bm{a}_i=-\frac{G M(<r_i)}{(r_i^2+\epsilon^2)^{3/2}}\bm{x}_i,
\label{eq:appforce}
\end{equation}
and integrated with a second-order kick--drift--kick scheme.

The dark matter and stellar components are initialized in equilibrium through Eddington inversion of their combined potential. Both are represented by Dehnen profiles~\cite{Dehnen:1993uh},
\begin{equation}
\rho(r)=\frac{(3-\gamma)Ma}{4\pi r^\gamma(r+a)^{4-\gamma}},
\label{eq:dehnen}
\end{equation}
fitted respectively to the digitized TNG100 dark matter profiles and observed stellar profiles of Ref.~\cite{Lei2026}. The stellar inner slope is fixed to $\gamma_\star=1$. Table~\ref{tab:ic} lists the fitted parameters and particle numbers. Each galaxy contains $10^6$ dark matter particles and $3.33\times10^5$ stellar particles.

\begin{table}[t]
 \caption{Initial-condition parameters for the four galaxy bins. $M$, $a$, and $\gamma$ denote the Dehnen parameters in Eq.~\eqref{eq:dehnen}. Each galaxy uses $N_{\rm DM}=10^6$ and $\gamma_\star=1$.}
 \label{tab:ic}
 \begin{ruledtabular}
  \begin{tabular}{lccccc}
   bin & $M_{\rm DM}$ & $a_{\rm DM}$ & $\gamma_{\rm DM}$
       & $M_\star$ & $a_\star$ \\
       & ($10^{11}M_\odot$) & (kpc) & & ($10^{9}M_\odot$) & (kpc)  \\
   \colrule
   1 & 2.872 & 65.9 & 1.694 & 5.98 & 1.17 \\
   2 & 8.502 & 117.5 & 1.828 & 22.8 & 3.00 \\
   3 & 30.54 & 181.6 & 1.703 & 75.6 & 3.63 \\
   4 & 277.9 & 480.4 & 1.565 & 229 & 6.08 \\
  \end{tabular}
 \end{ruledtabular}
\end{table}

The stellar component is live and responds to the evolving monopole. Positive-energy particles are retained, so halo expansion and evaporation arise dynamically. The two dark matter states have the same gravitational macro-particle mass, since corrections of order $\delta/m_\chi$ are negligible, while the internal-energy difference is retained explicitly in the collision kinematics. The spherical treatment assumes isotropy and does not capture rotation, triaxiality, substructure, or cosmological infall.

\subsection{Collisions}
\label{app:collisions}

Dark matter particles are grouped in radial cells and randomly paired at each collision step. The effective partner density is
\begin{equation}
\rho_{\rm pair}=
\frac{N(N-1)}{2\lfloor N/2\rfloor}
\frac{m_p}{V_{\rm cell}},
\label{eq:pairweight}
\end{equation}
so that sampling disjoint pairs reproduces the full unordered-pair collision rate in expectation. For a channel $c$ with relative speed $v_r$,
\begin{equation}
P_c=\rho_{\rm pair}K_c(v_r)\Delta t_{\rm col},
\qquad
K_c(v_r)\equiv\frac{\sigma_c(v_r)}{m_\chi}v_r.
\label{eq:appprobability}
\end{equation}
The collision interval is adaptively subdivided to keep the probability per sampled pair small.

The microscopic calculation supplies five channels,
\begin{equation}
22\rightarrow11,\quad
22\rightarrow22,\quad
11\rightarrow11,\quad
12\rightarrow12,\quad
11\rightarrow22,
\label{eq:appchannels}
\end{equation}
with tabulated velocity-dependent rates and angular distributions. Accepted channels are drawn according to their relative rates, and scattering angles are sampled from the corresponding differential cross sections.

Elastic events preserve the relative speed. Inelastic events obey
\begin{align}
v_{r,f}^2&=v_{r,i}^2+4v_{\rm k}^2 &&(22\rightarrow11),\nonumber\\
v_{r,f}^2&=v_{r,i}^2-4v_{\rm k}^2 &&(11\rightarrow22),
\label{eq:appkinematics}
\end{align}
so up-scattering is forbidden below $v_{r,i}=2v_k$. Pair momentum is conserved in the center-of-momentum frame, while the kinetic energy changes by the corresponding internal-energy release or absorption.

Within the spherical reduction, the unresolved tangential orientation is randomized before and after each collision. This preserves kinetic energy, radial transport, angular kinetic support, and the scattering-angle distribution statistically, while excluding coherent angular-momentum structures.

\subsection{Elastic validation}
\label{app:yy22}

We validate the collision algorithm in the single-component elastic limit using the BM2 halo of Yang and Yu~\cite{Yang:2022hkm}. The halo has
\begin{align}
 \rho_s&=2.74\times10^8\,M_\odot\,\mathrm{kpc}^{-3},\nonumber\\
 r_s&=0.141\,\mathrm{kpc},\qquad r_{200}=2.778\,\mathrm{kpc}.
 \label{eq:appbm2}
\end{align}
and is evolved with a constant isotropic cross section
$\sigma/m_\chi=7.1\,\cmg$, matching the calibration of Ref.~\cite{Yang:2023jwn}. The diagnostic is the mean density within $r_c=0.03\,\mathrm{kpc}$,
\begin{equation}
\bar\rho_c(t)=\frac{3M(<r_c,t)}{4\pi r_c^3}.
\label{eq:appcentraldensity}
\end{equation}

Because a finite particle realization requires an outer boundary, we initialize a smoothly truncated NFW profile,
\begin{equation}
\rho(r)=
\frac{\rho_s}{(r/r_s)(1+r/r_s)^2}
\exp[-(r/r_t)^2],
\qquad
r_t=2r_{200},
\label{eq:apptruncnfw}
\end{equation}
using its isotropic Eddington distribution function. The precise gravothermal timescale depends mildly on this outer truncation, which acts as the halo heat reservoir.

\begin{figure}[t]
\centering
\includegraphics[width=\columnwidth]{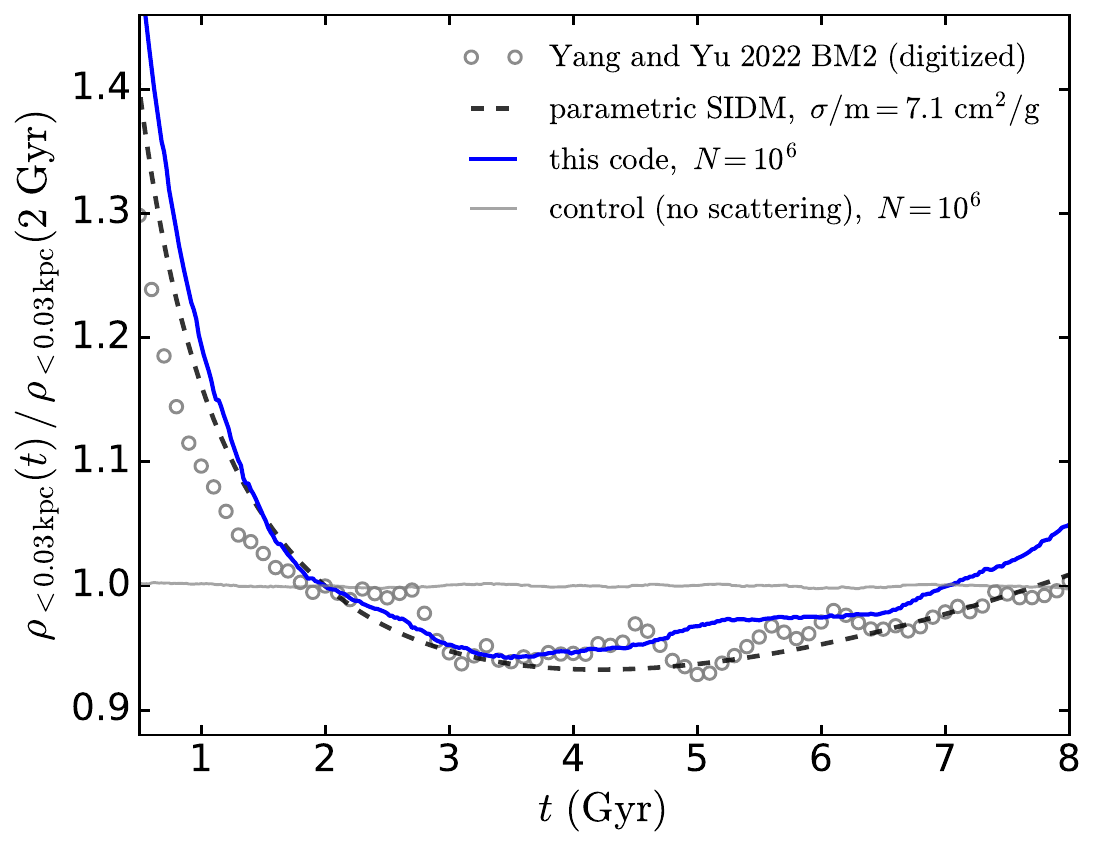}
\caption{Single-component elastic validation using the BM2 halo of Ref.~\cite{Yang:2022hkm} with $\sigma/m_\chi=7.1\,\cmg$. The blue curve shows the present $N=10^6$ calculation, hollow points the published simulation, the dashed curve the parametric model of Ref.~\cite{Yang:2023jwn}, and the gray curve a matched collisionless control. The calculation reproduces the core expansion, minimum density, and gravothermal recovery at the few-percent level.}
\label{fig:yy22validation}
\end{figure}

Figure~\ref{fig:yy22validation} compares our $N=10^6$ calculation with the published $N$-body evolution and the parametric gravothermal model of Ref.~\cite{Yang:2023jwn}. The calculation reproduces the initial core expansion, the density minimum, and the subsequent gravothermal recovery to within a few percent over $0.5-10\,\mathrm{Gyr}$. A matched collisionless run remains close to its initial density, confirming that the evolution is driven by collisional heat transport rather than numerical relaxation.

Production runs conserve energy to better than $10^{-5}$ and require no clipping of collision probabilities. The validation establishes the collision rate, center-of-momentum update, and gravothermal heat transport within the spherical particle treatment. The inelastic channels are additionally constrained by the event-by-event thresholds and energy conservation of Eq.~\eqref{eq:appkinematics}.

\end{document}